\documentstyle[11pt,newpasp,twoside,epsf]{article} 
\def\edcomment#1{\iffalse\marginpar{\raggedright\sl#1\/}\else\relax\fi} 
\reversemarginpar 

\begin{document} 
\title{Dynamical Evolution of Globular Clusters in the Magellanic Clouds and Other Local Systems}

\author{A. D. Mackey, G. F. Gilmore} 
\affil{Institute of Astronomy, Madingley Road, Cambridge, CB3 0HA, UK} 

\begin{abstract} 
Observations of rich LMC and SMC clusters reveal an increasing spread in core radius ($r_c$) with age. This trend likely represents real physical evolution in these systems. Old clusters appear either large or small. Similar $r_c$ distributions are observed for the Fornax and Sagittarius dSph cluster systems and in the ``young'' halo Milky Way subsystem. These observations have implications for the formation of the Galaxy and the dynamical evolution of globular clusters.
\end{abstract}

\section{Globular Clusters in the Magellanic Clouds} 

The Large and Small Magellanic Clouds (LMC/SMC) possess rich stellar
clusters of masses comparable to Milky Way globular clusters (MWGC), but
with ages $10^{6} \le \tau \le 10^{10}$ yr. These two systems therefore
permit direct observational studies of globular cluster evolution. We
have assembled and reduced archival Hubble Space Telescope (HST) data for 
53 LMC and 10 SMC clusters spanning the full age range, and have measured 
parameters such as the core radius $(r_c)$ and the total mass and 
luminosity for each cluster using surface brightness
profiles (Mackey \& Gilmore 2002a; 2002b). Our results, Figure 1(a), 
show a clear trend in $r_c$ with increasing age for both systems, 
including an apparent bifurcation of the distribution at 
$\tau \sim 10^{8}$ yr so that old clusters may have either large or 
small $r_c$. As discussed in Mackey \& Gilmore (2002a,
2002b), we conclude that the observed trend with age represents genuine 
physical evolution of clusters in these two systems. The mechanism by
which some clusters undergo large scale core expansion while otherwise
similar clusters do not is as yet unidentified, and we are exploring
possibilities via $N$-body simulations.

\section{Globular Clusters in nearby dSph galaxies and the Milky Way} 

\begin{figure}
\plottwo{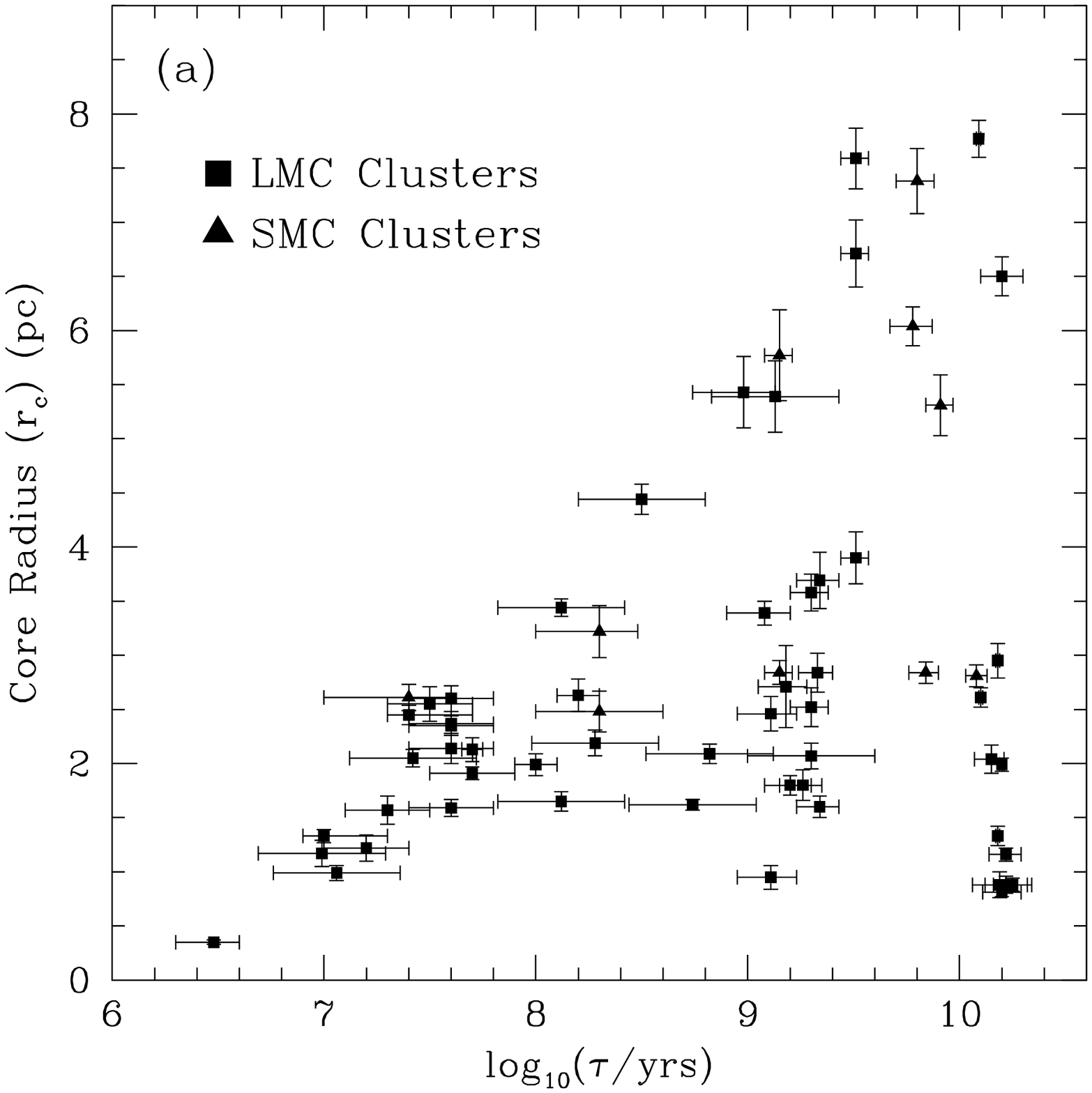}{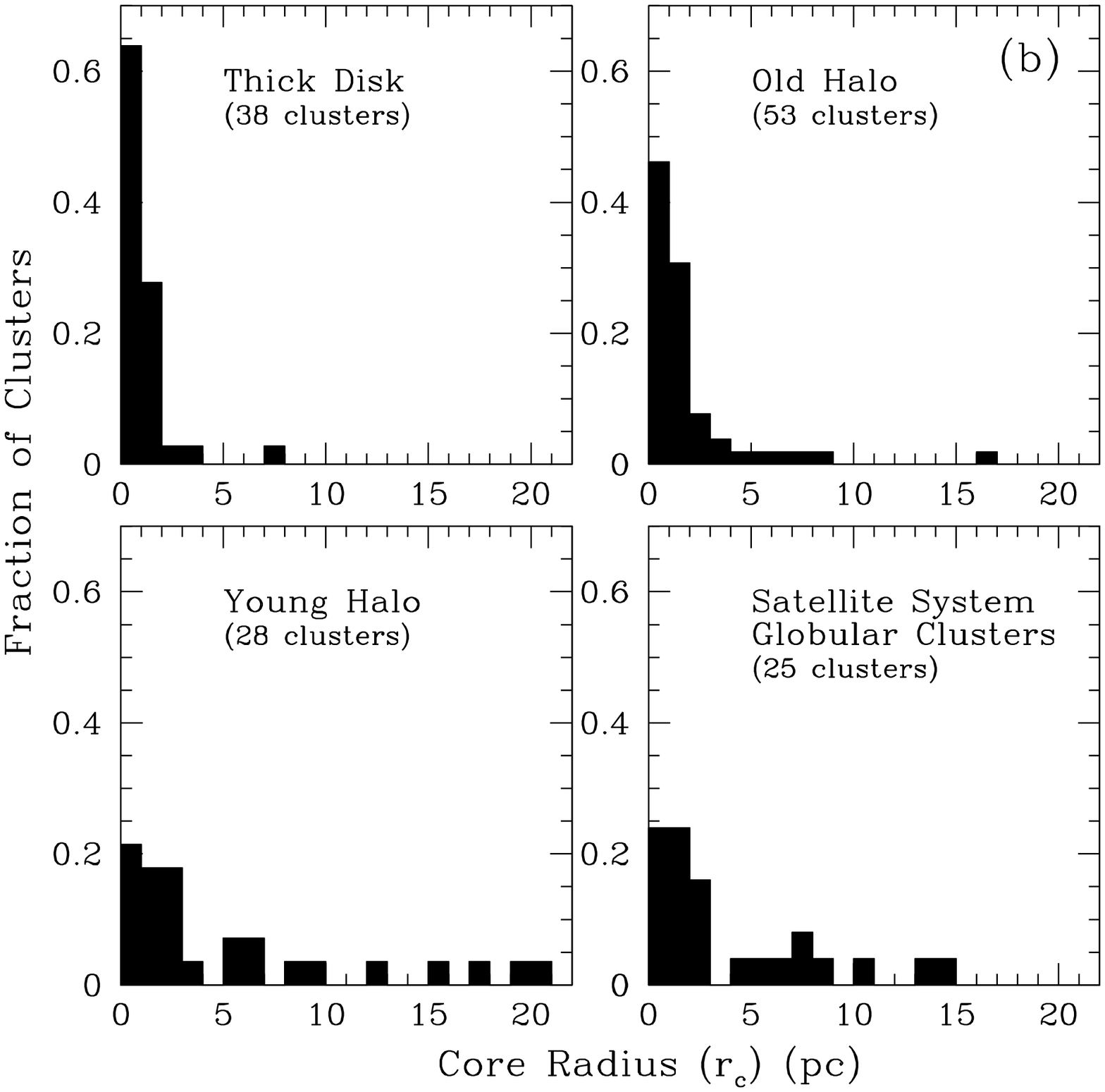}
\caption{(a) $r_c$ vs. age for our combined LMC/SMC sample. (b) Binned $r_c$ distributions for the three MWGC subsystems of Zinn (1993) and the combined ``satellite'' system.}
\end{figure}

Another avenue of investigation is to observe whether any other local
systems show globular clusters with expanded cores. Unlike the LMC/SMC
clusters, all other local globular clusters are old 
($\tau \sim 10^{10}$ yr). Fornax and Sagittarius (Sgr)
are the two most massive dSph galaxies associated with the Milky Way, and
the only two to have globular clusters. Fornax has five (C1-C5) while
Sgr has four (M54, Ter 7, Ter 8, Arp 2) plus one clearly 
previously associated (Pal 12). Again using archival HST data, we have 
measured $r_c$ of $10.0$ pc, $5.8$ pc, $1.6$ pc, $1.8$ pc, and 
$< 1.4$ pc for Fornax C1-C5 respectively (Mackey \& Gilmore 2002c). Like 
the old LMC/SMC clusters, these clusters are either compact 
($r_c \leq 3.5$ pc) or expanded ($r_c \geq 5.5$ pc). From the catalogue 
of Harris (1996) the Sgr clusters have $r_c$ 
of $0.9$ pc, $4.1$ pc, $7.6$ pc, $13.2$ pc, and $1.1$ pc for M54, Ter 7, 
Ter 8, Arp 2, and Pal 12 respectively. We again observe a similar $r_c$
distribution. We also examine the MWGC system, using the data
of Harris. The MWGC system may be split by [Fe/H] and horizontal branch
type according to the prescription of Zinn (1993). Plotting 
binned $r_c$ distributions for each subsystem, Figure 1(b), we
see each is distinct, with the ``young'' halo subsystem containing
the majority of clusters with large cores. We can form a composite
``satellite'' system, containing the old LMC/SMC clusters, and the 
Fornax and Sgr dSph clusters. The $r_c$ distribution for this system
appears very similar to that for the ``young'' halo subsystem. A simple 
K-S test shows the unbinned distributions to be similar at $\sim 93 \%$
significance. The satellite system and the thick-disk and ``old'' halo 
subsystems are different at $99.9\%$ and $99.8\%$ significances 
respectively. This result supports Zinn's (1993) hypothesis that the 
``young'' halo clusters have been accreted from destroyed satellite 
galaxies, while the thick-disk and ``old'' halo clusters are mostly 
intrinsic to the Milky Way. Alternatively, if one already accepts this 
picture, our result shows that clusters with large cores do not survive 
(or perhaps form) in regions with strong tidal fields. Core expansion 
may therefore be exclusively a weak-field effect.


\begin{references}
\reference Harris, W. E. 1996, \aj, 112, 1487
\reference Mackey, A. D., \& Gilmore, G. F. 2002a, \mnras, in press (astro-ph/0209031)
\reference Mackey, A. D., \& Gilmore, G. F. 2002b, \mnras, in press (astro-ph/0209046)
\reference Mackey, A. D., \& Gilmore, G. F. 2002c, \mnras, submitted
\reference Zinn, R. 1993, in ASP Conf. Ser. Vol. 48, The Globular Cluster-Galaxy Connection, ed. G. H. Smith \& J. P. Brodie (San Francisco: ASP), 303
\end{references}
\end{document}